\documentclass{llncs}

\usepackage{epsfig}
\usepackage{url}

\usepackage{tabularx}
\usepackage{multirow}
\usepackage{url}
\usepackage{booktabs}

\usepackage{graphicx}
\usepackage{caption}

\usepackage{subcaption}
\usepackage{cite}

\begin{document}

\title{Enriching Ontologies with Encyclopedic Background Knowledge for Document Indexing}

\author{Lisa Posch\inst{}}

\institute{GESIS -- Leibniz Institute for the Social Sciences\\
Unter Sachsenhausen 6-8 \\
D-50667 Cologne, Germany \\
\email{lisa.posch@gesis.org}}

\maketitle

\begin{abstract}

The rapidly increasing number of scientific documents available publicly on the Internet creates the challenge of efficiently organizing and indexing these documents.
Due to the time consuming and tedious nature of manual classification and indexing, there is a need for better methods to automate this process. 
This thesis proposes an approach which leverages encyclopedic background knowledge for enriching domain-specific ontologies with textual and structural information about the semantic vicinity of the ontologies' concepts.
The proposed approach aims to exploit this information for improving both ontology-based methods for classifying and indexing documents and methods based on supervised machine learning.

\end{abstract}

\subsubsection*{Acknowledgments.}
The thesis presented in this paper is supervised by Prof. Dr. Markus Strohmaier.

\section{Introduction}

The amount of scientific publications available on the Internet is increasing rapidly. Without efficient methods for document classification and indexing, it is increasingly time consuming and difficult for researchers to find relevant publications.  
Traditionally, scientific institutions performed the task of facilitating search for relevant literature by manually indexing and classifying new publications, with the goal of maintaining an ideally complete domain-specific database. 
However, this task is becoming progressively more difficult to perform, as manual indexing is time consuming, tedious and expensive. 

In the recent decades, researchers of different domains have attempted to tackle this problem by developing a wide range of methods for automatic text classification and indexing. Most of these methods are based on machine learning algorithms or on algorithms which use ontologies as background knowledge. 
While existing approaches allow rapid classification and indexing of a large number of documents, the quality of the results is not comparable to the performance of expert human indexers. Therefore, it is an ongoing challenge to improve the methods for automatic classification and indexing. 

The main goal of this thesis is to build upon existing methods to construct an improved framework for automatic classification and subject indexing of documents. The proposed approach leverages encyclopedic background knowledge for enriching existing domain-specific ontologies and classification systems with additional textual and structural information about the semantic vicinity of the ontologies' concepts.
Specifically, I plan to investigate whether this encyclopedic background knowledge is useful for improving the results of ontology-based classification and indexing methods as well as methods based on machine learning. 

This paper is structured as follows: Section \ref{relatedwork} gives a brief overview of related research on subject indexing and classification. My research questions and hypotheses are stated in Section \ref{researchquestions}. Section \ref{approach} discusses the proposed approach and Section \ref{evaluation} describes the evaluation process. The datasets which I plan to apply my approach to are introduced in Section \ref{datasets}; preliminary experiments on one of these datasets are presented in Section \ref{preliminaryresults}. Finally, Section \ref{conclusion} concludes this work.

\section{Related Work}
\label{relatedwork}

Most of the approaches to classification and indexing of documents are based on either machine learning algorithms or on methods which use ontologies as background knowledge. 
This section briefly summarizes the main techniques which have been used to address the challenge of automatically classifying and indexing documents.

\emph{Subject indexing} refers to assigning topical keywords to documents (usually from a controlled  vocabulary such as a thesaurus), 
while \emph{document classification} assigns a document to one or more semantic categories. The difference between these tasks, however, is negligible, as the aim of both is to produce appropriate connections between documents and semantic entities \cite{lancaster2003indexing}.
\emph{Semantic annotation} refers to attaching meta-data to resources, usually in the context of the Semantic Web \cite{oren2006semantic}. 
Both subject indexing and document classification can therefore be seen as a form of semantic document annotation.

\textbf{Machine Learning Based Methods:} 
Both supervised and unsupervised machine learning methods have been applied to document classification and indexing.
A popular method for representing documents in supervised learning is the \emph{bag-of-words} approach, which represents documents by the words they contain, disregarding the order in which they occur. Instead of single words, also sequences of words (\emph{n-grams}) can be used to represent a document. The words or n-grams can be weighted by different schemes such as term frequency or TF-IDF \cite{salton1988term}.
While TF-IDF is unable to capture the semantic structures in documents, methods such as \emph{latent semantic analysis (LSA)} \cite{deerwester1990indexing} and \emph{probabilistic LSA (pLSA)} \cite{hofmann1999probabilistic} try to overcome this weakness. 
Recently, also encyclopedic background knowledge has been leveraged for representing documents. For example, \emph{explicit semantic analysis (ESA)} \cite{gabrilovich2007computing} represents the meaning of documents as weighted vectors of Wikipedia-based concepts. While originally intended for computing semantic relatedness, it has also been applied successfully to text classification (e.g. \cite{gupta2008text}).

One commonly used method for unsupervised text classification and indexing are \emph{topic models}. Topic models are statistical models which aim to discover latent topics in documents. The simplest one is \emph{Latent Dirichlet Allocation (LDA)}, which was introduced by Blei et al. \cite{blei2003latent}. 
A supervised version of LDA, \emph{sLDA}, was later presented by Blei and McAuliffe \cite{blei2007supervised}.
\emph{Labeled LDA}, another supervised topic model, was introduced by Ramage et al. \cite{ramage2009labeled}. Labeled LDA constrains the latent topics which are to be learned to the labels of the documents in the training dataset.
Topic models have also been used to create features for training supervised classifiers \cite{blei2007supervised}.

\textbf{Ontology-Based Methods:} 
An ontology is a \emph{``formal, explicit specification of a shared conceptualization."} \cite{studer1998knowledge}
Ontologies have been used as background knowledge for semantic annotation of documents (e.g. \cite{jonquet2009open}, \cite{de2007multilingual}), mostly in the context of the Semantic Web. 
Jonquet et al. \cite{jonquet2009open} presented the Open Biomedical Annotator, which is an ontology-based Web service for annotating documents with biomedical ontology concepts. The annotation process consists of two main steps: The first step, \emph{concept recognition} produces direct annotations by matching textual meta-data of the documents to ontology concepts. In the second step, the set of direct annotations is \emph{expanded} by using semantic relations of the ontology and by using existing mappings to other ontologies. 
I plan to build upon and extend this approach by incorporating textual and structural encyclopedic background knowledge.

\section{Research Questions and Hypotheses}
\label{researchquestions}

My research aims to investigate ways in which encyclopedic background knowledge, in the form of textual and structural information about the semantic vicinity of ontology concepts, may be useful for improving the classification and indexing of documents. 
In particular, I plan to address the following research questions:

\begin{enumerate}
\item How does the effectiveness of automatic indexing and classification techniques which exploit encyclopedic background knowledge compare to the effectiveness of techniques which do not use encyclopedic background knowledge?
\item How does the effectiveness of automatic indexing and classification techniques which exploit encyclopedic background knowledge change with different strategies of incorporating the background knowledge?
\item In which ways can encyclopedic background knowledge be useful for effectively combining the sets of keywords and classes suggested by machine learning methods with those suggested by ontology-based methods?
\end{enumerate}

I hypothesize that encyclopedic background knowledge about the semantic neighborhood of the defined concepts in an ontology can be successfully leveraged for modeling a more comprehensive representation of said concepts and that the enhanced representation of these concepts is likely to contribute to a more accurate classification and indexing of documents.

\section{Proposed Approach}
\label{approach}

This section describes the proposed approach for classifying and indexing documents using encyclopedic background knowledge.
The approach leverages non-domain-specific encyclopedic background knowledge to enrich existing domain-specific ontologies (and classification systems) with additional information about the concepts contained in the ontology. This additional information includes encyclopedic textual information about concepts which are semantically closely related to concepts contained in the ontology, as well as structural information about the nature of the semantic relations between the concepts.
I believe that this information can be useful for automatic classification and indexing of documents, contributing to a more accurate assignment of semantic classes and topical keywords by ontology-based as well as supervised methods.

To the best of my knowledge, this is the first attempt to enrich existing domain-specific ontologies and classification systems with non-domain-specific encyclopedic background knowledge with the aim of improving automatic indexing and classification of documents. 
The two main steps of my approach are described in more detail in the rest of this section.

\subsection{Enriching an Existing Domain-Specific Ontology}

The first step of the approach consists in enriching an existing domain-specific ontology with encyclopedic background knowledge.
This can be achieved by first mapping the concepts contained in the ontology to the concepts contained in an encyclopedia and subsequently modeling the semantic neighborhood of the ontology's concepts.
Wikipedia constitutes an attractive option for using as the encyclopedia of choice, as 
it has often shown to be useful for a wide range of applications in domains such as natural language processing, information retrieval and ontology building \cite{medelyan2009mining}. The mapping could be achieved either manually or by employing automatic mapping techniques. 

For modeling the semantic neighborhood of the ontology's concepts, it is necessary to identify which encyclopedic entries lie in the semantic vicinity of the ontology's concepts, as well as the nature of the semantic relations to the ontology's concepts and between the encyclopedic entries. If Wikipedia is chosen as the encyclopedia, this task can be achieved by employing ontologies extracted from Wikipedia such as Yago \cite{suchanek2007yago} or DBpedia \cite{auer2007dbpedia}. 
One of the main challenges in this step is to adequately map the entire ontology, and to appropriately deal with ontology concepts which do not match any encyclopedia concept (e.g., by linking them to several related encyclopedia concepts).

\subsection{Using the Enriched Ontology with Existing Classification and Indexing Methods}

The second step of the approach is to investigate whether the enriched ontology is useful for classifying and indexing documents. To achieve this, I plan to integrate it into existing ontology-based methods and supervised machine learning methods. 

\textbf{Ontology-Based Methods:} 
In an approach building upon the one used by the Open Biomedical Annotator \cite{jonquet2009open}, 
encyclopedic background knowledge is likely to be useful in both concept recognition and the identification of semantically related ontology concepts to extend the keyword set. Concerning the concept recognition task, the textual information from encyclopedic entries in semantic vicinity of ontology concepts can be used to better identify matching concepts in the text (i.a. by alleviating the problem of vocabulary mismatch). Regarding the extension of the keyword set, 
the structural information from the encyclopedia can provide support nodes and support relations, 
which could prove useful for making a better decision on which additional ontology concepts to include in the keyword set.

\textbf{Supervised Machine Learning Methods:} 
The textual information of the encyclopedic articles in the semantic vicinity of a category or ontology concept is likely to be able to diminish the problem of sparseness  in training datasets, by providing additional training examples. A potential limitation for the usefulness of these additional training examples is that the type of language used in the documents may differ significantly from the language used in the encyclopedia.

\textbf{Combination of Ontology-Based and Supervised Methods:} 
Encyclopedic background knowledge could be useful for effectively combining the keywords suggested by supervised machine learning with those suggested by the ontology-based approach. The resulting keyword sets could be combined, for example, by starting out with the intersection set and including further keywords from the union of both keyword sets. Which keywords are chosen for indexing would depend on the semantic distance, calculated on the enriched ontology, to the keywords in the intersection set. 

\section{Evaluation}
\label{evaluation}

To evaluate the utility of the enriched ontology for the various methods, I plan to use three methods which I describe in this section. Each method will be employed to compare indexing and classification methods which use the enriched ontology with the corresponding methods which do not use the enriched ontology.

\textbf{Evaluation based on existing manually created keyword sets:} The results of the different models will be compared to existing manually defined keywords provided by expert indexers. 
Standard metrics such as precision, recall and F1-Score can be used to quantify the effectiveness of the models.

\textbf{Comparison with inter-expert semantic similarity:} Semantic similarity measures can be employed for calculating the semantic distance between different sets of suggested keywords. The semantic distance between the keyword sets produced by different human expert indexers (for the same document) will be measured. This semantic distance will then be compared to the semantic distance between the keyword set produced by the model and the keyword sets produced by the different human expert indexers. The choice of the semantic similarity measure can be based on a preceding evaluation of the accuracy of the results of different measures, conducted by domain experts.

\textbf{Recommendation-based evaluation by expert indexers:} The keyword sets produced by the model will be presented to human expert indexers, along with the document to be indexed. The human annotators will then judge which suggested keywords, in their opinion, are appropriate, 
which ones are wrong and which ones are missing. Based on this evaluation, standard metrics can be calculated, as well as the semantic distance to the keyword set after correction. 

\begin{figure}[t!]
        \centering
        \begin{subfigure}[b]{0.39\textwidth}
                \includegraphics[width=\textwidth]{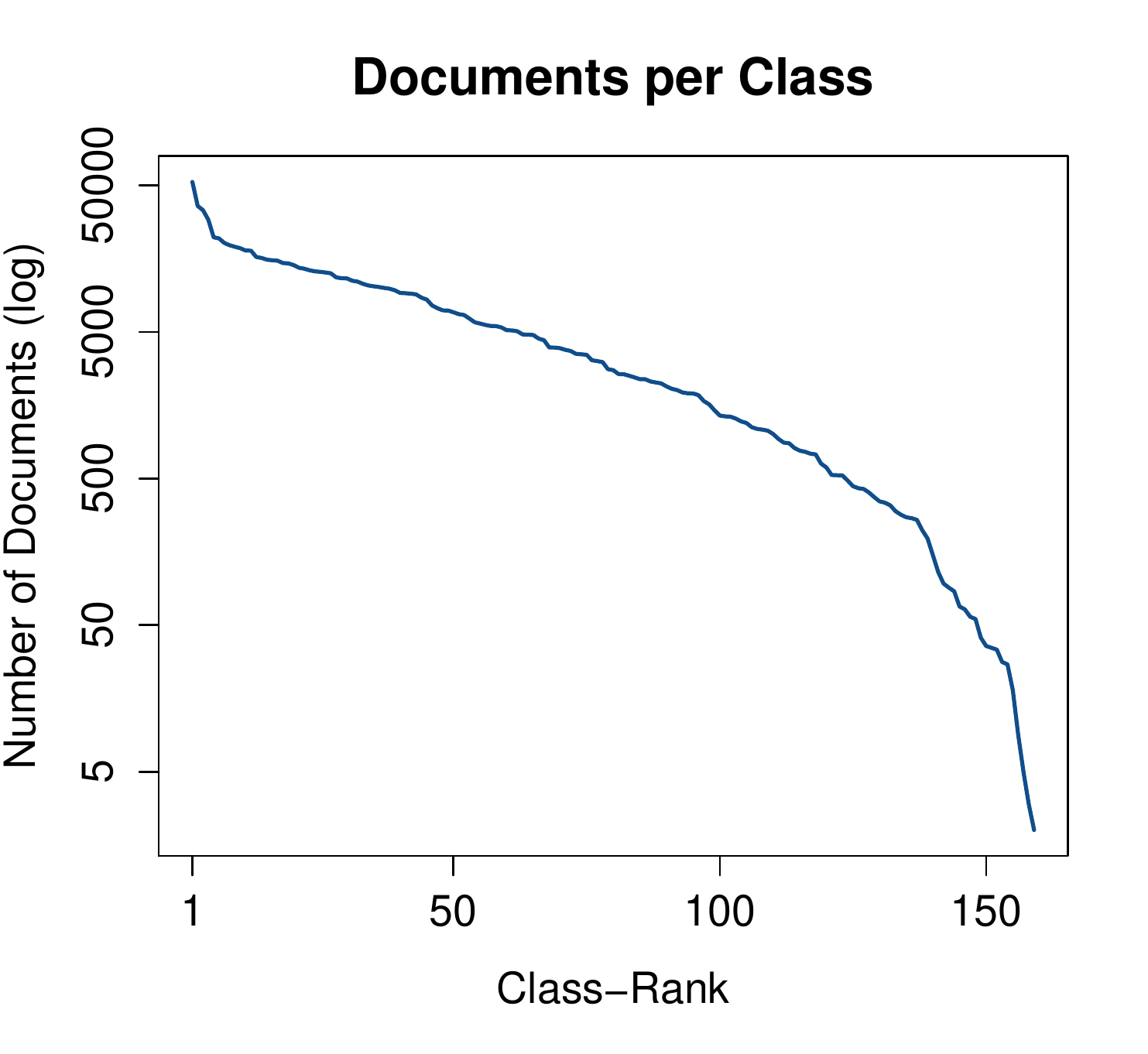}
                \label{fig:tfidf}
        \end{subfigure}%
        ~ 
        \begin{subfigure}[b]{0.39\textwidth}
                \includegraphics[width=\textwidth]{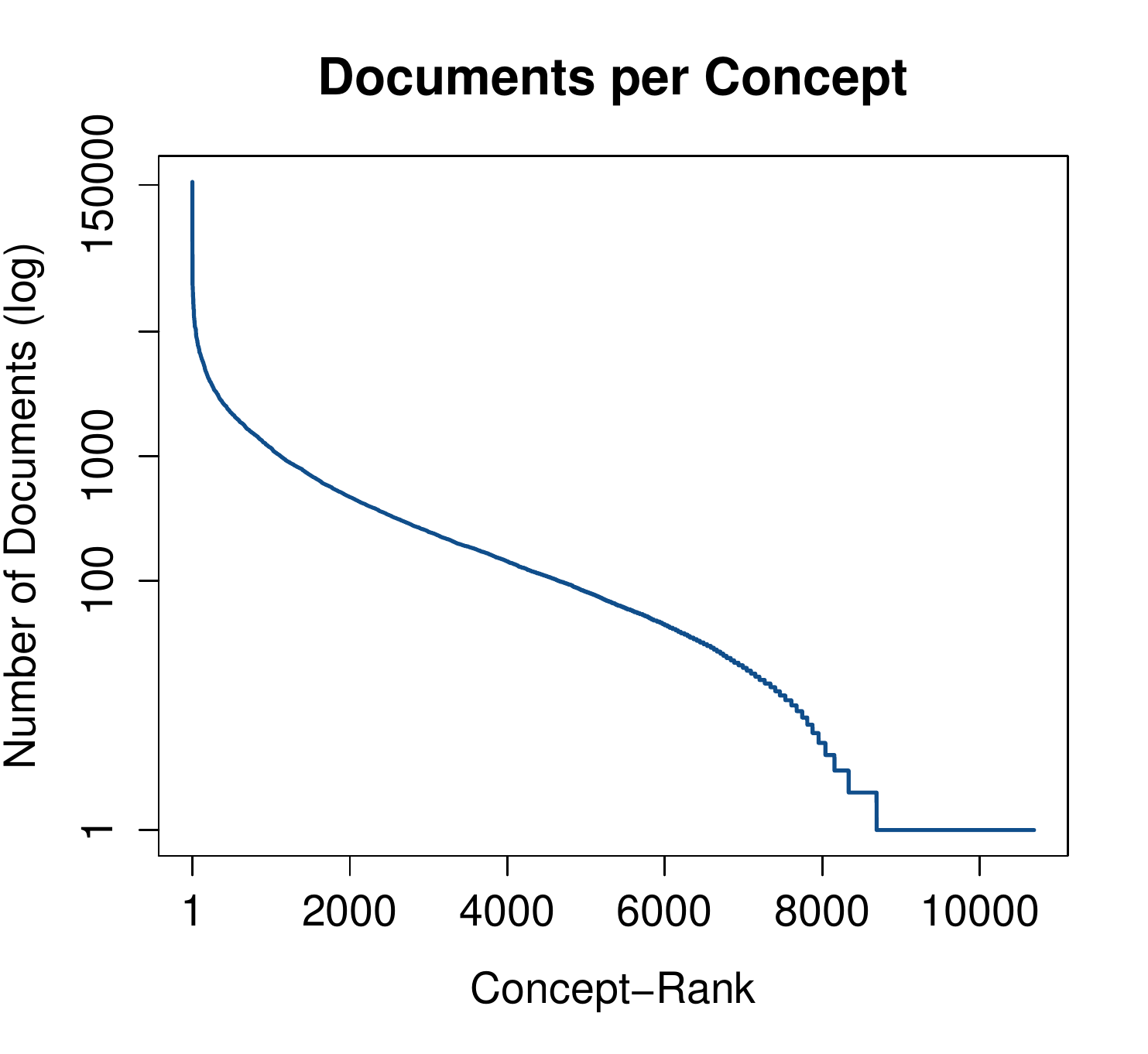}
                \label{fig:llda}
        \end{subfigure}
				\vspace{-5pt}
				\caption{Distribution of documents (log-scale) in the different classes from the classification system for the social sciences and concepts from the thesaurus for the social sciences in the SOLIS dataset.} 
				\label{fig:distribution}
\end{figure} 

\section{Datasets}
\label{datasets}

I plan to apply my approach to and evaluate it on the following datasets: the Social Science Literature Information System (SOLIS), the Social Science Open Access Repository (databases containing German social science publications), the German Education Index (German educational science publications), and PubMed Central (English biomedical and life sciences publications). 
The first dataset which I will apply my approach to is SOLIS, a collection of meta-data (including abstracts) of roughly 450,000 social science publications which are fully manually classified and indexed by human expert indexers according to the classification system and the thesaurus for the social sciences.

\section{Preliminary Results}
\label{preliminaryresults}

This section briefly describes the supervised classification experiments which I conducted on the SOLIS dataset. The results presented in this section are not to be seen as preliminary results of my proposed approach, but rather as a motivation for why an improved approach is necessary.


\textbf{Experimental Setup:}  
I conducted the classification experiments on a subset of the SOLIS database which consists of all documents that were published after the year 2003. It contains 144,259 documents and 306,879 class labels (from the classification system for the social sciences). 
Figure \ref{fig:distribution} shows the skewed distribution of classes and ontology concepts assigned to documents. A skewed class distribution is often a problem when applying supervised classification methods, due to the lack of training documents in the sparse classes.


After applying standard preprocessing techniques (removing stopwords and stemming), I calculated two sets of features on the textual information of the documents: \emph{TF-IDF features} and \emph{Labeled LDA topic distributions}. 
Identifying the categories assigned to a document constitutes a multi-label classification task, where there can be multiple correct classes for each document. I used the \emph{One-vs-Rest} strategy for this task, which trains a separate classifier for each class and fits this class against all other classes.
The classification system for the social sciences is hierarchically organized, so it is possible to conduct classification at different levels of semantic specificity.
Three classification models, one for each of the three top semantic levels in the classification hierarchy, were trained for both feature sets. Support vector machines with linear kernels were used for all classification experiments, and all classifiers were trained on a 67\% split of the subset and tested on the remaining 33\%.

\begin{figure}[t!]
        \centering
        \begin{subfigure}[b]{0.4\textwidth}
                \includegraphics[width=\textwidth]{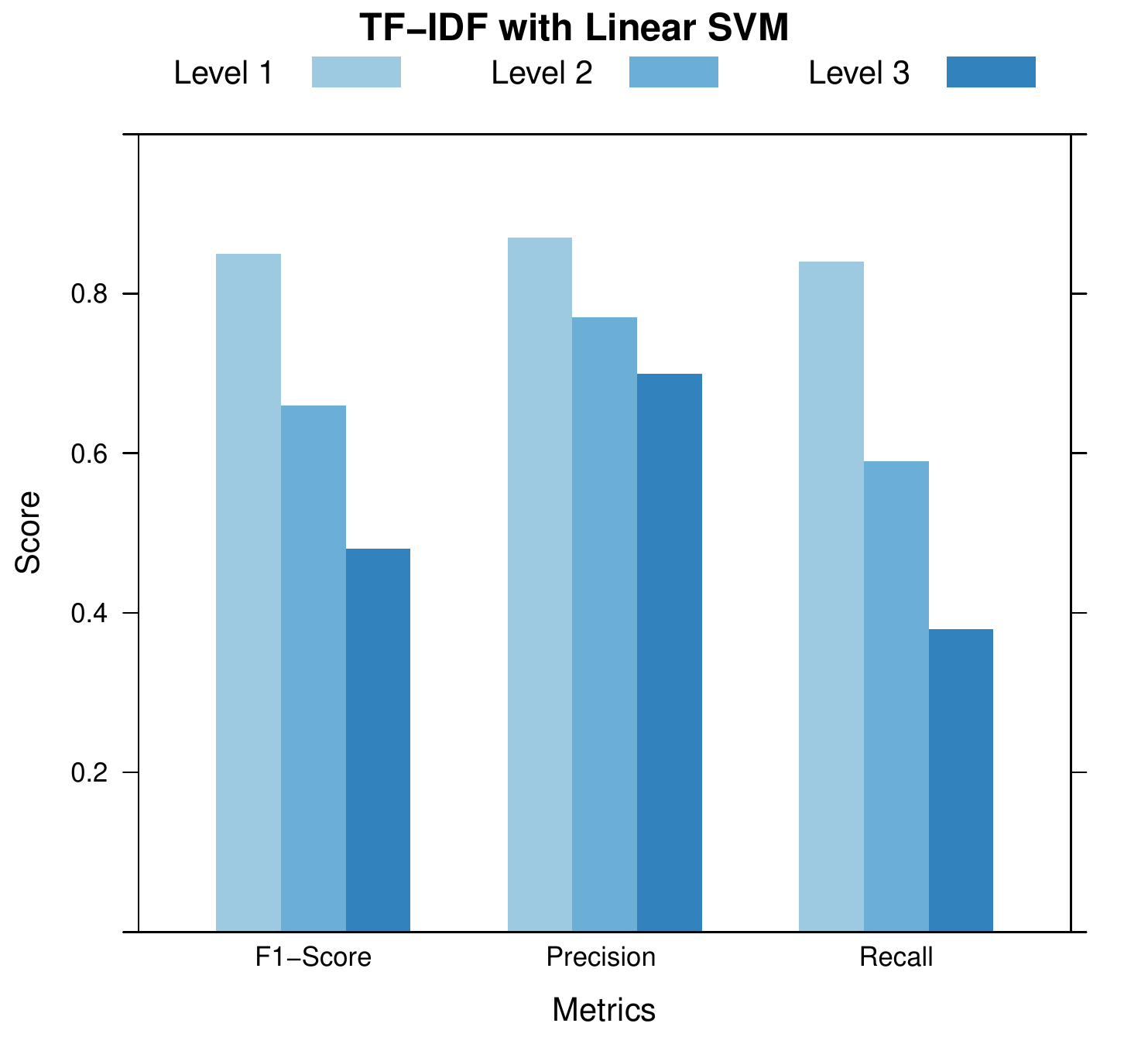}
                \caption{TF-IDF with linear SVM.}
                \label{fig:tfidf}
        \end{subfigure}%
        ~ 
        \begin{subfigure}[b]{0.4\textwidth}
                \includegraphics[width=\textwidth]{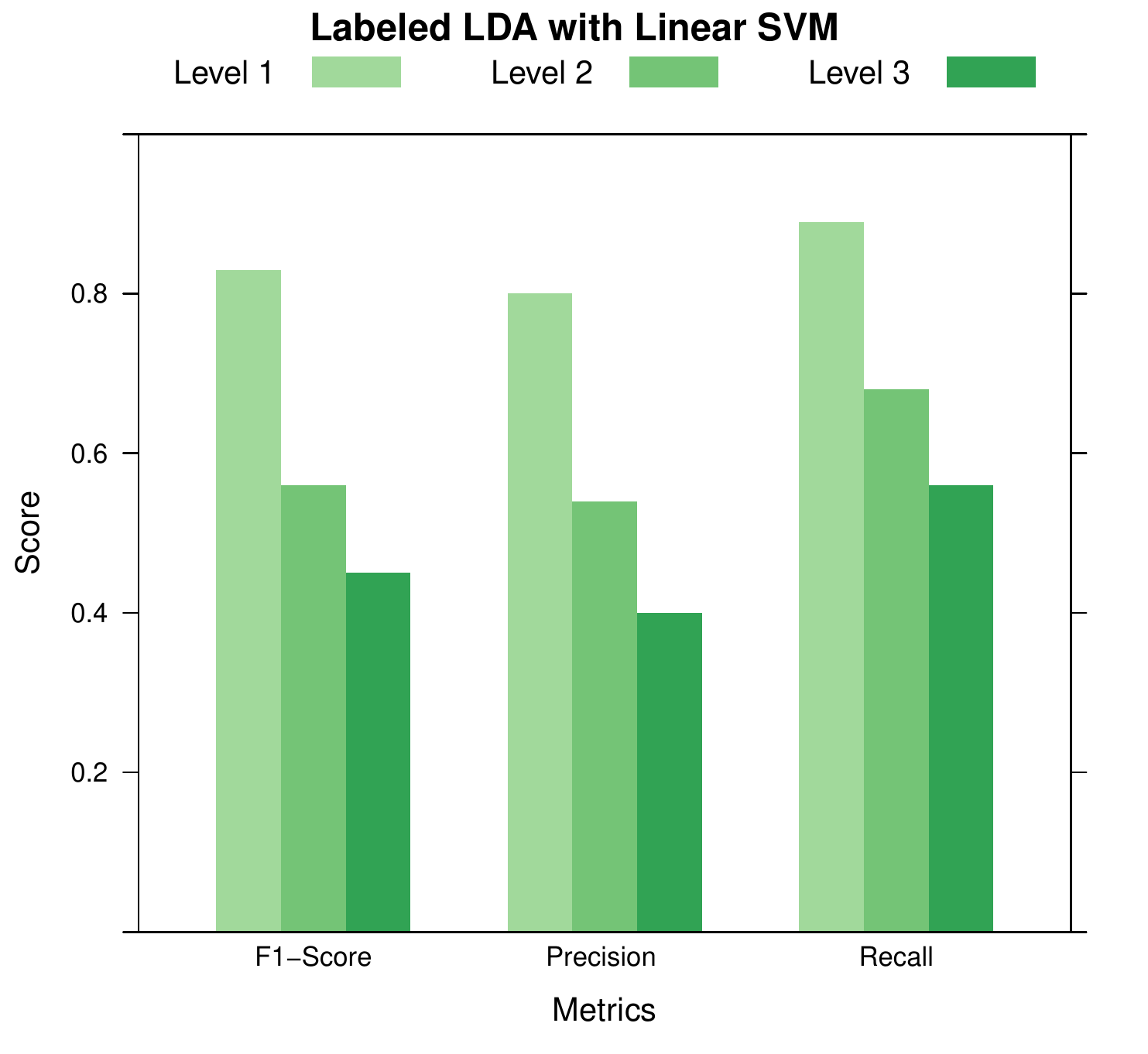}
                \caption{Labeled LDA with linear SVM.}
                \label{fig:llda}
        \end{subfigure}
				\caption{Results of supervised machine learning targeting three semantic specificity levels of the classification system.}\label{fig:results}
\end{figure}

\textbf{Results and Discussion:} 
The results of the experiments are presented in Figure \ref{fig:results}. Generally, Labeled LDA features produced a higher recall, while TF-IDF features resulted in a higher precision. While both feature sets achieve an acceptable F1-Score when targeting only first-level hierarchy categories, the effectiveness of the classifiers targeting more semantically fine-grained categories is unsatisfactory. This shows that a more elaborate approach is necessary for effectively classifying social science documents.

\section{Conclusion}
\label{conclusion}

Subject indexing of unstructured text continues to constitute a challenging field of research. The PhD thesis presented in this paper focuses on a new approach to enriching ontologies and classification systems with the background knowledge of encyclopedias. From such background knowledge, textual and structural information about the semantic vicinity of ontologies' concepts can be extracted.
This additional knowledge, by providing a more comprehensive representation of concepts contained in an ontology, is likely to be useful for automatically indexing and classifying documents.

\bibliographystyle{splncs03}
\bibliography{biblio_consortium}

\end{document}